
%
%
%
\def\unredoffs{} \def\redoffs{\voffset=-.31truein\hoffset=-.59truein}
\def\speclscape{\special{ps: landscape}}
%
%
%
%
\newbox\leftpage \newdimen\fullhsize \newdimen\hstitle \newdimen\hsbody
\tolerance=1000\hfuzz=2pt
\catcode`\@=11 
\def\bigans{b }
\def\answ{b }
\ifx\answ\bigans\message{(This will come out unreduced.}
\magnification=1200\unredoffs\baselineskip=16pt plus 2pt minus 1pt
\hsbody=\hsize \hstitle=\hsize 
\else\message{(This will be reduced.} \let\l@r=L
\magnification=1000\baselineskip=16pt plus 2pt minus 1pt \vsize=7truein
\redoffs \hstitle=8truein\hsbody=4.75truein\fullhsize=10truein\hsize=\hsbody
\output={\ifnum\pageno=0 
  \shipout\vbox{\speclscape{\hsize\fullhsize\makeheadline}
    \hbox to \fullhsize{\hfill\pagebody\hfill}}\advancepageno
  \else
  \almostshipout{\leftline{\vbox{\pagebody\makefootline}}}\advancepageno
  \fi}
\def\almostshipout#1{\if L\l@r \count1=1 \message{[\the\count0.\the\count1]}
      \global\setbox\leftpage=#1 \global\let\l@r=R
 \else \count1=2
  \shipout\vbox{\speclscape{\hsize\fullhsize\makeheadline}
      \hbox to\fullhsize{\box\leftpage\hfil#1}}  \global\let\l@r=L\fi}
\fi
%
\newcount\yearltd\yearltd=\year\advance\yearltd by -1900

\def\Title#1#2{\nopagenumbers\abstractfont\hsize=\hstitle\rightline{#1}%
\vskip 1in\centerline{\titlefont #2}\abstractfont\vskip .5in\pageno=0}
\def\Date#1{\vfill\leftline{#1}\tenpoint\supereject\global\hsize=\hsbody%
\footline={\hss\tenrm\folio\hss}}
%

\def\draftmode{\message{ DRAFTMODE }\def\draftdate{{\rm preliminary draft:
\number\month/\number\day/\number\yearltd\ \ \hourmin}}%
\headline={\hfil\draftdate}\writelabels\baselineskip=20pt plus 2pt minus 2pt
 {\count255=\time\divide\count255 by 60 \xdef\hourmin{\number\count255}
  \multiply\count255 by-60\advance\count255 by\time
  \xdef\hourmin{\hourmin:\ifnum\count255<10 0\fi\the\count255}}}
\def\nolabels{\def\wrlabeL##1{}\def\eqlabeL##1{}\def\reflabeL##1{}}
\def\writelabels{\def\wrlabeL##1{\leavevmode\vadjust{\rlap{\smash%
{\line{{\escapechar=` \hfill\rlap{\sevenrm\hskip.03in\string##1}}}}}}}%
\def\eqlabeL##1{{\escapechar-1\rlap{\sevenrm\hskip.05in\string##1}}}%
\def\reflabeL##1{\noexpand\llap{\noexpand\sevenrm\string\string\string##1}}}
\nolabels
%
\global\newcount\secno \global\secno=0
\global\newcount\meqno \global\meqno=1
\def\newsec#1{\global\advance\secno by1\message{(\the\secno. #1)}
\global\subsecno=0\eqnres@t\noindent{\bf\the\secno. #1}
\writetoca{{\secsym} {#1}}\par\nobreak\medskip\nobreak}
\def\eqnres@t{\xdef\secsym{\the\secno.}\global\meqno=1\bigbreak\bigskip}
\def\sequentialequations{\def\eqnres@t{\bigbreak}}\xdef\secsym{}
\global\newcount\subsecno \global\subsecno=0
\def\subsec#1{\global\advance\subsecno by1\message{(\secsym\the\subsecno.
#1)}
\ifnum\lastpenalty>9000\else\bigbreak\fi
\noindent{\it\secsym\the\subsecno. #1}\writetoca{\string\quad
{\secsym\the\subsecno.} {#1}}\par\nobreak\medskip\nobreak}
\def\appendix#1#2{\global\meqno=1\global\subsecno=0\xdef\secsym{\hbox{#1.}}
\bigbreak\bigskip\noindent{\bf Appendix #1. #2}\message{(#1. #2)}
\writetoca{Appendix {#1.} {#2}}\par\nobreak\medskip\nobreak}
%
%
\def\eqnn#1{\xdef #1{(\secsym\the\meqno)}\writedef{#1\leftbracket#1}%
\global\advance\meqno by1\wrlabeL#1}
\def\eqna#1{\xdef #1##1{\hbox{$(\secsym\the\meqno##1)$}}
\writedef{#1\numbersign1\leftbracket#1{\numbersign1}}%
\global\advance\meqno by1\wrlabeL{#1$\{\}$}}
\def\eqn#1#2{\xdef #1{(\secsym\the\meqno)}\writedef{#1\leftbracket#1}%
\global\advance\meqno by1$$#2\eqno#1\eqlabeL#1$$}
%
\newskip\footskip\footskip14pt plus 1pt minus 1pt 
\def\footnotefont{\ninepoint}\def\f@t#1{\footnotefont #1\@foot}
\def\f@@t{\baselineskip\footskip\bgroup\footnotefont\aftergroup\@foot\let\next}
\setbox\strutbox=\hbox{\vrule height9.5pt depth4.5pt width0pt}
\global\newcount\ftno \global\ftno=0
\def\foot{\global\advance\ftno by1\footnote{$^{\the\ftno}$}}
%
\newwrite\ftfile
\def\footend{\def\foot{\global\advance\ftno by1\chardef\wfile=\ftfile
$^{\the\ftno}$\ifnum\ftno=1\immediate\openout\ftfile=foots.tmp\fi%
\immediate\write\ftfile{\noexpand\smallskip%
\noexpand\item{f\the\ftno:\ }\pctsign}\findarg}%
\def\footatend{\vfill\eject\immediate\closeout\ftfile{\parindent=20pt
\centerline{\bf Footnotes}\nobreak\bigskip\input foots.tmp }}}
\def\footatend{}
%
%
\global\newcount\refno \global\refno=1
\newwrite\rfile
\def\ref{[\the\refno]\nref}
\def\nref#1{\xdef#1{[\the\refno]}\writedef{#1\leftbracket#1}%
\ifnum\refno=1\immediate\openout\rfile=refs.tmp\fi
\global\advance\refno by1\chardef\wfile=\rfile\immediate
\write\rfile{\noexpand\item{#1\ }\reflabeL{#1\hskip.31in}\pctsign}\findarg}
\def\findarg#1#{\begingroup\obeylines\newlinechar=`\^^M\pass@rg}
{\obeylines\gdef\pass@rg#1{\writ@line\relax #1^^M\hbox{}^^M}%
\gdef\writ@line#1^^M{\expandafter\toks0\expandafter{\striprel@x #1}%
\edef\next{\the\toks0}\ifx\next\em@rk\let\next=\endgroup\else\ifx\next\empty%
\else\immediate\write\wfile{\the\toks0}\fi\let\next=\writ@line\fi\next\relax}}
\def\striprel@x#1{} \def\em@rk{\hbox{}}
\def\lref{\begingroup\obeylines\lr@f}
\def\lr@f#1#2{\gdef#1{\ref#1{#2}}\endgroup\unskip}
\def\semi{;\hfil\break}
\def\addref#1{\immediate\write\rfile{\noexpand\item{}#1}} 
\def\startrefs#1{\immediate\openout\rfile=refs.tmp\refno=#1}
\def\xref{\expandafter\xr@f}\def\xr@f[#1]{#1}
\def\refs#1{\count255=1[\r@fs #1{\hbox{}}]}
\def\r@fs#1{\ifx\und@fined#1\message{reflabel \string#1 is undefined.}%
\nref#1{need to supply reference \string#1.}\fi%
\vphantom{\hphantom{#1}}\edef\next{#1}\ifx\next\em@rk\def\next{}%
\else\ifx\next#1\ifodd\count255\relax\xref#1\count255=0\fi%
\else#1\count255=1\fi\let\next=\r@fs\fi\next}
%

%
\newwrite\ffile\global\newcount\figno \global\figno=1
\def\fig{fig.~\the\figno\nfig}
\def\nfig#1{\xdef#1{fig.~\the\figno}%
\writedef{#1\leftbracket fig.\noexpand~\the\figno}%
\ifnum\figno=1\immediate\openout\ffile=figs.tmp\fi\chardef\wfile=\ffile%
\immediate\write\ffile{\noexpand\medskip\noexpand\item{Fig.\ \the\figno. }
\reflabeL{#1\hskip.55in}\pctsign}\global\advance\figno by1\findarg}
\def\xfig{\expandafter\xf@g}\def\xf@g fig.\penalty\@M\ {}
\def\figs#1{figs.~\f@gs #1{\hbox{}}}
\def\f@gs#1{\edef\next{#1}\ifx\next\em@rk\def\next{}\else
\ifx\next#1\xfig #1\else#1\fi\let\next=\f@gs\fi\next}
\newwrite\lfile
{\escapechar-1\xdef\pctsign{\string\%}\xdef\leftbracket{\string\{}
\xdef\rightbracket{\string\}}\xdef\numbersign{\string\#}}

\def\writestop{\def\writestoppt{\immediate\write\lfile{\string\pageno%
\the\pageno\string\startrefs\leftbracket\the\refno\rightbracket%
\string\def\string\secsym\leftbracket\secsym\rightbracket%
\string\secno\the\secno\string\meqno\the\meqno}\immediate\closeout\lfile}}
\def\writestoppt{}\def\writedef#1{}
\def\seclab#1{\xdef #1{\the\secno}\writedef{#1\leftbracket#1}\wrlabeL{#1=#1}}
\def\subseclab#1{\xdef #1{\secsym\the\subsecno}%
\writedef{#1\leftbracket#1}\wrlabeL{#1=#1}}
\newwrite\tfile \def\writetoca#1{}
\def\leaderfill{\leaders\hbox to 1em{\hss.\hss}\hfill}
\def\writetoc{\immediate\openout\tfile=toc.tmp
   \def\writetoca##1{{\edef\next{\write\tfile{\noindent ##1
   \string\leaderfill {\noexpand\number\pageno} \par}}\next}}}

%
\catcode`\@=12 
%
\edef\tfontsize{\ifx\answ\bigans scaled\magstep3\else scaled\magstep4\fi}
\font\titlerm=cmr10 \tfontsize \font\titlerms=cmr7 \tfontsize
\font\titlermss=cmr5 \tfontsize \font\titlei=cmmi10 \tfontsize
\font\titleis=cmmi7 \tfontsize \font\titleiss=cmmi5 \tfontsize
\font\titlesy=cmsy10 \tfontsize \font\titlesys=cmsy7 \tfontsize
\font\titlesyss=cmsy5 \tfontsize \font\titleit=cmti10 \tfontsize
\skewchar\titlei='177 \skewchar\titleis='177 \skewchar\titleiss='177
\skewchar\titlesy='60 \skewchar\titlesys='60 \skewchar\titlesyss='60
\def\titlefont{\def\rm{\fam0\titlerm}
\textfont0=\titlerm \scriptfont0=\titlerms \scriptscriptfont0=\titlermss
\textfont1=\titlei \scriptfont1=\titleis \scriptscriptfont1=\titleiss
\textfont2=\titlesy \scriptfont2=\titlesys \scriptscriptfont2=\titlesyss
\textfont\itfam=\titleit \def\it{\fam\itfam\titleit}\rm}
 \ifx\answ\bigans\else scaled\magstep1\fi
\ifx\answ\bigans\def\abstractfont{\tenpoint}\else
\font\abssl=cmsl10 scaled \magstep1
\font\absrm=cmr10 scaled\magstep1 \font\absrms=cmr7 scaled\magstep1
\font\absrmss=cmr5 scaled\magstep1 \font\absi=cmmi10 scaled\magstep1
\font\absis=cmmi7 scaled\magstep1 \font\absiss=cmmi5 scaled\magstep1
\font\abssy=cmsy10 scaled\magstep1 \font\abssys=cmsy7 scaled\magstep1
\font\abssyss=cmsy5 scaled\magstep1 \font\absbf=cmbx10 scaled\magstep1
\skewchar\absi='177 \skewchar\absis='177 \skewchar\absiss='177
\skewchar\abssy='60 \skewchar\abssys='60 \skewchar\abssyss='60
\def\abstractfont{\def\rm{\fam0\absrm}
\textfont0=\absrm \scriptfont0=\absrms \scriptscriptfont0=\absrmss
\textfont1=\absi \scriptfont1=\absis \scriptscriptfont1=\absiss
\textfont2=\abssy \scriptfont2=\abssys \scriptscriptfont2=\abssyss
\textfont\itfam=\bigit \def\it{\fam\itfam\bigit}\def\footnotefont{\tenpoint}%
\textfont\slfam=\abssl \def\sl{\fam\slfam\abssl}%
\textfont\bffam=\absbf \def\bf{\fam\bffam\absbf}\rm}\fi
\def\tenpoint{\def\rm{\fam0\tenrm}
\textfont0=\tenrm \scriptfont0=\sevenrm \scriptscriptfont0=\fiverm
\textfont1=\teni  \scriptfont1=\seveni  \scriptscriptfont1=\fivei
\textfont2=\tensy \scriptfont2=\sevensy \scriptscriptfont2=\fivesy
\textfont\itfam=\tenit
\def\it{\fam\itfam\tenit}\def\footnotefont{\ninepoint}%
\textfont\bffam=\tenbf \def\bf{\fam\bffam\tenbf}\def\sl{\fam\slfam\tensl}\rm}
\font\ninerm=cmr9 \font\sixrm=cmr6 \font\ninei=cmmi9 \font\sixi=cmmi6
\font\ninesy=cmsy9 \font\sixsy=cmsy6 \font\ninebf=cmbx9
\font\nineit=cmti9 \font\ninesl=cmsl9 \skewchar\ninei='177
\skewchar\sixi='177 \skewchar\ninesy='60 \skewchar\sixsy='60
\def\ninepoint{\def\rm{\fam0\ninerm}
\textfont0=\ninerm \scriptfont0=\sixrm \scriptscriptfont0=\fiverm
\textfont1=\ninei \scriptfont1=\sixi \scriptscriptfont1=\fivei
\textfont2=\ninesy \scriptfont2=\sixsy \scriptscriptfont2=\fivesy
\textfont\itfam=\ninei \def\it{\fam\itfam\nineit}\def\sl{\fam\slfam\ninesl}%
\textfont\bffam=\ninebf \def\bf{\fam\bffam\ninebf}\rm}
%
%

\hyphenation{anom-aly anom-alies coun-ter-term coun-ter-terms}
\def\inv{^{\raise.15ex\hbox{${\scriptscriptstyle -}$}\kern-.05em 1}}

\def\Dsl{\,\raise.15ex\hbox{/}\mkern-13.5mu D} 
\def\dsl{\raise.15ex\hbox{/}\kern-.57em\partial}

\font\bigit=cmti10 scaled \magstep1
\def\lspace{\ifx\answ\bigans{}\else\qquad\fi}
\def\lbspace{\ifx\answ\bigans{}\else\hskip-.2in\fi} 
\def\boxeqn#1{\vcenter{\vbox{\hrule\hbox{\vrule\kern3pt\vbox{\kern3pt
           \hbox{${\displaystyle #1}$}\kern3pt}\kern3pt\vrule}\hrule}}}
\def\mbox#1#2{\vcenter{\hrule \hbox{\vrule height#2in
               \kern#1in \vrule} \hrule}}  
%

\def\darr#1{\raise1.5ex\hbox{$\leftrightarrow$}\mkern-16.5mu #1}

\def\roughly#1{\raise.3ex\hbox{$#1$\kern-.75em\lower1ex\hbox{$\sim$}}}



\def\IB{\relax\hbox{$\inbar\kern-.3em{\rm B}$}}
\def\IC{\relax\hbox{$\inbar\kern-.3em{\rm C}$}}
\def\ID{\relax\hbox{$\inbar\kern-.3em{\rm D}$}}
\def\IE{\relax\hbox{$\inbar\kern-.3em{\rm E}$}}
\def\IF{\relax\hbox{$\inbar\kern-.3em{\rm F}$}}
\def\IG{\relax\hbox{$\inbar\kern-.3em{\rm G}$}}
\def\IGa{\relax\hbox{${\rm I}\kern-.18em\Gamma$}}
\def\IH{\relax{\rm I\kern-.18em H}}
\def\IK{\relax{\rm I\kern-.18em K}}
\def\II{\relax{\rm I\kern-.18em I}}
\def\IL{\relax{\rm I\kern-.18em L}}
\def\IP{\relax{\rm I\kern-.18em P}}
\def\IR{\relax{\rm I\kern-.18em R}}
\def\IZ{\relax\ifmmode\mathchoice {\hbox{\cmss Z\kern-.4em Z}}{\hbox{\cmss
Z\kern-.4em Z}} {\lower.9pt\hbox{\cmsss Z\kern-.4em Z}}
{\lower1.2pt\hbox{\cmsss Z\kern-.4em Z}}\else{\cmss Z\kern-.4em Z}\fi}

\def\IB{\relax{\rm I\kern-.18em B}}
\def\IC{{\relax\hbox{$\inbar\kern-.3em{\rm C}$}}}
\def\ID{\relax{\rm I\kern-.18em D}}
\def\IE{\relax{\rm I\kern-.18em E}}
\def\IF{\relax{\rm I\kern-.18em F}}


\def\CW {{\cal W}}

\def\p{\partial}




\def\s{\lies}


\def\c{\cdot}


\def\f{\phi}    
    
\def\F{\Phi}

\def\a{\alpha}
\def\b{\beta}
  
\def\d{\delta}  \def\D{\Delta}
\def\m{\mu}

\def\r{\rho}
\def\l{\lambda} \def\L{\Lambda}

\def\|{\Big|}
\def\({\Big(}   \def\){\Big)}
\def\[{\Big[}   \def\]{\Big]}



\def\paper#1#2#3#4{#1, {\sl #2}, #3 {\tt #4}}

\def\hh{hep-th/}


\def\PLB#1#2#3{Phys. Lett.~{\bf B#1} (#2) #3}
\def\NPB#1#2#3{Nucl. Phys.~{\bf B#1} (#2) #3}
\def\PRL#1#2#3{Phys. Rev. Lett.~{\bf #1} (#2) #3}
\def\CMP#1#2#3{Comm. Math. Phys.~{\bf #1} (#2) #3}
\def\PRD#1#2#3{Phys. Rev.~{\bf D#1} (#2) #3}
\def\MPL#1#2#3{Mod. Phys. Lett.~{\bf #1} (#2) #3}
\def\IJMP#1#2#3{Int. Jour. Mod. Phys.~{\bf #1} (#2) #3}


\def\unlockat{\catcode`\@=11}
\def\lockat{\catcode`\@=12}

\unlockat


\def\newsec#1{\global\advance\secno by1\message{(\the\secno. #1)}
\global\subsecno=0\global\subsubsecno=0\eqnres@t\noindent {\bf\the\secno. #1}
\writetoca{{\secsym} {#1}}\par\nobreak\medskip\nobreak}
\global\newcount\subsecno \global\subsecno=0
\def\subsec#1{\global\advance\subsecno by1\message{(\secsym\the\subsecno.
#1)}
\ifnum\lastpenalty>9000\else\bigbreak\fi\global\subsubsecno=0
\noindent{\it\secsym\the\subsecno. #1}
\writetoca{\string\quad {\secsym\the\subsecno.} {#1}}
\par\nobreak\medskip\nobreak}
\global\newcount\subsubsecno \global\subsubsecno=0
\def\subsubsec#1{\global\advance\subsubsecno by1
\message{(\secsym\the\subsecno.\the\subsubsecno. #1)}
\ifnum\lastpenalty>9000\else\bigbreak\fi
\noindent\quad{\secsym\the\subsecno.\the\subsubsecno.}{#1}
\writetoca{\string\qquad{\secsym\the\subsecno.\the\subsubsecno.}{#1}}
\par\nobreak\medskip\nobreak}

\def\subsubseclab#1{\DefWarn#1\xdef #1{\noexpand\hyperref{}{subsubsection}%
{\secsym\the\subsecno.\the\subsubsecno}%
{\secsym\the\subsecno.\the\subsubsecno}}%
\writedef{#1\leftbracket#1}\wrlabeL{#1=#1}}
\lockat

\def\dbend{\lower3.5pt\hbox{\manual\char127}}


\def\boxit#1{\vbox{\hrule\hbox{\vrule\kern8pt
\vbox{\hbox{\kern8pt}\hbox{\vbox{#1}}\hbox{\kern8pt}}
\kern8pt\vrule}\hrule}}

\def\mathboxit#1{\vbox{\hrule\hbox{\vrule\kern8pt\vbox{\kern8pt
\hbox{$\displaystyle #1$}\kern8pt}\kern8pt\vrule}\hrule}}


\def\inbar{\,\vrule height1.5ex width.4pt depth0pt}

\font\cmss=cmss10 \font\cmsss=cmss10 at 7pt


\lref\simons{ J. Cheeger and J. Simons, {\it Differential Characters and
Geometric Invariants},  Stony Brook Preprint, (1973), unpublished.}

\lref\cargese{ L.~Baulieu, {\it Algebraic quantization of gauge theories},
Perspectives in fields and particles, Plenum Press, eds. Basdevant-Levy,
Cargese Lectures 1983}

\lref\antifields{ L. Baulieu, M. Bellon, S. Ouvry, C.Wallet, Phys.Letters
B252 (1990) 387; M.  Bocchichio, Phys. Lett. B187 (1987) 322;  Phys. Lett. B
192 (1987) 31; R.  Thorn    Nucl. Phys.   B257 (1987) 61. }

\lref\thompson{ George Thompson,  Annals Phys. 205 (1991) 130; J.M.F.
Labastida, M. Pernici, Phys. Lett. 212B  (1988) 56; D. Birmingham, M.Blau,
M. Rakowski and G.Thompson, Phys. Rept. 209 (1991) 129.}

\lref\tonin{ Tonin}

\lref\wittensix{ E.  Witten, {\it New  Gauge  Theories In Six Dimensions},
\hh{9710065}. }

\lref\orlando{ O. Alvarez, L. A. Ferreira and J. Sanchez Guillen, {\it  A New
Approach to Integrable Theories in any Dimension}, hep-th/9710147.}

\lref\wittentopo{ E.  Witten,  {\it  Topological Quantum Field Theory},
\hh9403195, Commun.  Math. Phys.  {117} (1988)353.  }

\lref\wittentwist{ E.  Witten, {\it Supersymmetric Yang--Mills theory on a
four-manifold}, J.  Math.  Phys.  {35} (1994) 5101.}

\lref\west{ L.~Baulieu, P.~West, {\it Six Dimensional TQFTs and  Self-dual
Two-Forms,} Phys.Lett. B {\bf 436 } (1998) 97, /hep-th/9805200}

\lref\bv{ I.A. Batalin and V.A. Vilkowisky,    Phys. Rev.   D28  (1983)
2567\semi M. Henneaux,  Phys. Rep.  126   (1985) 1\semi M. Henneaux and C.
Teitelboim, {\it Quantization of Gauge Systems}
  Princeton University Press,  Princeton (1992).}

\lref\kyoto{ L. Baulieu, E. Bergschoeff and E. Sezgin, Nucl. Phys.
B307(1988)348\semi L. Baulieu,   {\it Field Antifield Duality, p-Form Gauge
Fields
   and Topological Quantum Field Theories}, hep-th/9512026,
   Nucl. Phys. B478 (1996) 431.  }

\lref\sourlas{ G. Parisi and N. Sourlas, {\it Random Magnetic Fields,
Supersymmetry and Negative Dimensions}, Phys. Rev. Lett.  43 (1979) 744;
Nucl.  Phys.  B206 (1982) 321.  }

\lref\SalamSezgin{ A.  Salam  and  E.  Sezgin, {\it Supergravities in
diverse dimensions}, vol.  1, p. 119\semi P.  Howe, G.  Sierra and P.
Townsend, Nucl Phys B221 (1983) 331.}

\lref\nekrasov{ A. Losev, G. Moore, N. Nekrasov, S. Shatashvili, {\it
Four-Dimensional Avatars of Two-Dimensional RCFT},  hep-th/9509151, Nucl.
Phys.  Proc.  Suppl.   46 (1996) 130\semi L.  Baulieu, A.  Losev,
N.~Nekrasov  {\it Chern-Simons and Twisted Supersymmetry in Higher
Dimensions},  hep-th/9707174, to appear in Nucl.  Phys.  B.  }

\lref\WitDonagi{R.~ Donagi, E.~ Witten, ``Supersymmetric Yang--Mills Theory
and Integrable Systems'', hep-th/9510101, Nucl. Phys.{\bf B}460 (1996)
299-334}
\lref\Witfeb{E.~ Witten, ``Supersymmetric Yang--Mills Theory On A
Four-Manifold,''  hep-th/9403195; J. Math. Phys. {\bf 35} (1994) 5101.}
\lref\Witgrav{E.~ Witten, ``Topological Gravity'', Phys.Lett.206B:601, 1988}
\lref\witaffl{I. ~ Affleck, J.A.~ Harvey and E.~ Witten,
        ``Instantons and (Super)Symmetry Breaking
        in $2+1$ Dimensions'', Nucl. Phys. {\bf B}206 (1982) 413}
\lref\wittabl{E.~ Witten,  ``On $S$-Duality in Abelian Gauge Theory,''
hep-th/9505186; Selecta Mathematica {\bf 1} (1995) 383}
\lref\wittgr{E.~ Witten, ``The Verlinde Algebra And The Cohomology Of The
Grassmannian'',  hep-th/9312104}
\lref\wittenwzw{E. Witten, ``Non abelian bosonization in two dimensions,''
Commun. Math. Phys. {\bf 92} (1984)455 }
\lref\witgrsm{E. Witten, ``Quantum field theory, grassmannians and algebraic
curves,'' Commun.Math.Phys.113:529,1988}
\lref\wittjones{E. Witten, ``Quantum field theory and the Jones
polynomial,'' Commun.  Math. Phys., 121 (1989) 351. }
\lref\witttft{E.~ Witten, ``Topological Quantum Field Theory", Commun. Math.
Phys. {\bf 117} (1988) 353.}
\lref\wittmon{E.~ Witten, ``Monopoles and Four-Manifolds'', hep-th/9411102}
\lref\Witdgt{ E.~ Witten, ``On Quantum gauge theories in two dimensions,''
Commun. Math. Phys. {\bf  141}  (1991) 153}
\lref\witrevis{E.~ Witten,
 ``Two-dimensional gauge theories revisited'', hep-th/9204083; J. Geom.
Phys. 9 (1992) 303-368}
\lref\Witgenus{E.~ Witten, ``Elliptic Genera and Quantum Field Theory'',
Comm. Math. Phys. 109(1987) 525. }
\lref\OldZT{E. Witten, ``New Issues in Manifolds of SU(3) Holonomy,'' {\it
Nucl. Phys.} {\bf B268} (1986) 79 \semi J. Distler and B. Greene, ``Aspects
of (2,0) String Compactifications,'' {\it Nucl. Phys.} {\bf B304} (1988) 1
\semi B. Greene, ``Superconformal Compactifications in Weighted Projective
Space,'' {\it Comm. Math. Phys.} {\bf 130} (1990) 335.}
\lref\bagger{E.~ Witten and J. Bagger, Phys. Lett. {\bf 115B}(1982) 202}
\lref\witcurrent{E.~ Witten,``Global Aspects of Current Algebra'',
Nucl.Phys.B223 (1983) 422\semi ``Current Algebra, Baryons and Quark
Confinement'', Nucl.Phys. B223 (1993) 433}
\lref\Wittreiman{S.B. Treiman, E. Witten, R. Jackiw, B. Zumino, ``Current
Algebra and Anomalies'', Singapore, Singapore: World Scientific ( 1985) }
\lref\Witgravanom{L. Alvarez-Gaume, E.~ Witten, ``Gravitational Anomalies'',
Nucl.Phys.B234:269,1984. }

\lref\nicolai{\paper {H.~Nicolai}{New Linear Systems for 2D Poincar\'e
Supergravities}{\NPB{414}{1994}{299},}{\hh 9309052}.}



\lref\bg{\paper {L.~Baulieu, B.~Grossman}{Monopoles and Topological Field
Theory}{\PLB{214}{1988}{223}.}{}}

\lref\seibergsix{\paper {N.~Seiberg}{Non-trivial Fixed Points of The
Renormalization Group in Six
 Dimensions}{\PLB{390}{1997}{169}}{\hh 9609161}\semi
\paper {O.J.~Ganor, D.R.~Morrison, N.~Seiberg}{
  Branes, Calabi-Yau Spaces, and Toroidal Compactification of the N=1
  Six-Dimensional $E_8$ Theory}{\NPB{487}{1997}{93}}{\hh 9610251}\semi
\paper {O.~Aharony, M.~Berkooz, N.~Seiberg}{Light-Cone
  Description of (2,0) Superconformal Theories in Six
  Dimensions}{Adv. Theor. Math. Phys. {\bf 2} (1998) 119}{\hh 9712117.}}

\lref\cs{\paper {L.~Baulieu}{Chern-Simons Three-Dimensional and
Yang--Mills-Higgs Two-Dimensional Systems as Four-Dimensional Topological
Quantum Field Theories}{\PLB{232}{1989}{473}.}{}}

\lref\beltrami{\paper {L.~Baulieu, M.~Bellon}{Beltrami Parametrization and
String Theory}{\PLB{196}{1987}{142}}{}\semi
\paper {L.~Baulieu, M.~Bellon, R.~Grimm}{Beltrami Parametrization For
Superstrings}{\PLB{198}{1987}{343}}{}\semi
\paper {R.~Grimm}{Left-Right Decomposition of Two-Dimensional Superspace
Geometry and Its BRS Structure}{Annals Phys. {\bf 200} (1990) 49.}{}}

\lref\bbg{\paper {L.~Baulieu, M.~Bellon, R.~Grimm}{Left-Right Asymmetric
Conformal Anomalies}{\PLB{228}{1989}{325}.}{}}

\lref\bonora{\paper {G.~Bonelli, L.~Bonora, F.~Nesti}{String Interactions
from Matrix String Theory}{\NPB{538}{1999}{100},}{\hh 9807232}\semi
\paper {G.~Bonelli, L.~Bonora, F.~Nesti, A.~Tomasiello}{Matrix String Theory
and its Moduli Space}{}{\hh 9901093.}}

\lref\corrigan{\paper {E.~Corrigan, C.~Devchand, D.B.~Fairlie,
J.~Nuyts}{First Order Equations for Gauge Fields in Spaces of Dimension
Greater Than Four}{\NPB{214}{452}{1983}.}{}}

\lref\acha{\paper {B.S.~Acharya, M.~O'Loughlin, B.~Spence}{Higher
Dimensional Analogues of Donaldson-Witten Theory}{\NPB{503}{1997}{657},}{\hh
9705138}\semi
\paper {B.S.~Acharya, J.M.~Figueroa-O'Farrill, M.~O'Loughlin,
B.~Spence}{Euclidean
  D-branes and Higher-Dimensional Gauge   Theory}{\NPB{514}{1998}{583},}{\hh
  9707118.}}

\lref\Witr{\paper{E.~Witten}{Introduction to Cohomological Field   Theories}
{Lectures at Workshop on Topological Methods in Physics (Trieste, Italy, Jun
11-25, 1990), \IJMP{A6}{1991}{2775}.}{}}

\lref\ohta{\paper {L.~Baulieu, N.~Ohta}{Worldsheets with Extended
Supersymmetry} {\PLB{391}{1997}{295},}{\hh 9609207}.}

\lref\gravity{\paper {L.~Baulieu}{Transmutation of Pure 2-D Supergravity
Into Topological 2-D Gravity and Other Conformal Theories}
{\PLB{288}{1992}{59},}{\hh 9206019.}}

\lref\wgravity{\paper {L.~Baulieu, M.~Bellon, R.~Grimm}{Some Remarks on  the
Gauging of the Virasoro and   $w_{1+\infty}$
Algebras}{\PLB{260}{1991}{63}.}{}}

\lref\fourd{\paper {E.~Witten}{Topological Quantum Field
Theory}{\CMP{117}{1988}{353}}{}\semi
\paper {L.~Baulieu, I.M.~Singer}{Topological Yang--Mills Symmetry}{Nucl.
Phys. Proc. Suppl. {\bf 15B} (1988) 12.}{}}

\lref\topo{\paper {L.~Baulieu}{On the Symmetries of Topological Quantum Field
Theories}{\IJMP{A10}{1995}{4483},}{\hh 9504015}\semi
\paper {R.~Dijkgraaf, G.~Moore}{Balanced Topological Field
Theories}{\CMP{185}{1997}{411},}{\hh 9608169.}}

\lref\wwgravity{\paper {I.~Bakas} {The Large $N$ Limit   of Extended
Conformal Symmetries}{\PLB{228}{1989}{57}.}{}}

\lref\wwwgravity{\paper {C.M.~Hull}{Lectures on $\CW$-Gravity,
$\CW$-Geometry and
$\CW$-Strings}{}{\hh 9302110}, and~references therein.}

\lref\wvgravity{\paper {A.~Bilal, V.~Fock, I.~Kogan}{On the origin of
$W$-algebras}{\NPB{359}{1991}{635}.}{}}

\lref\surprises{\paper {E.~Witten} {Surprises with Topological Field
Theories} {Lectures given at ``Strings 90'', Texas A\&M, 1990,}{Preprint
IASSNS-HEP-90/37.}}

\lref\stringsone{\paper {L.~Baulieu, M.B.~Green, E.~Rabinovici}{A Unifying
Topological Action for Heterotic and  Type II Superstring  Theories}
{\PLB{386}{1996}{91},}{\hh 9606080.}}

\lref\stringsN{\paper {L.~Baulieu, M.B.~Green, E.~Rabinovici}{Superstrings
from   Theories with $N>1$ World Sheet Supersymmetry}
{\NPB{498}{1997}{119},}{\hh 9611136.}}

\lref\bks{\paper {L.~Baulieu, H.~Kanno, I.~Singer}{Special Quantum Field
Theories in Eight and Other Dimensions}{\CMP{194}{1998}{149},}{\hh
9704167}\semi
\paper {L.~Baulieu, H.~Kanno, I.~Singer}{Cohomological Yang--Mills Theory
  in Eight Dimensions}{ Talk given at APCTP Winter School on Dualities in
String Theory (Sokcho, Korea, February 24-28, 1997),} {\hh 9705127.}}

\lref\witdyn{\paper {P.~Townsend}{The eleven dimensional supermembrane
revisited}{\PLB{350}{1995}{184},}{\hh9501068}\semi
\paper{E.~Witten}{String Theory Dynamics in Various Dimensions}
{\NPB{443}{1995}{85},}{\hh 9503124}.}

\lref\bfss{\paper {T.~Banks, W.Fischler, S.H.~Shenker,
L.~Susskind}{$M$-Theory as a Matrix Model~:
A~Conjecture}{\PRD{55}{1997}{5112},} {\hh9610043.}}

\lref\seiberg{\paper {N.~Seiberg}{Why is the Matrix Model
Correct?}{\PRL{79}{1997}{3577},} {\hh 9710009.}}

\lref\sen{\paper {A.~Sen}{$D0$ Branes on $T^n$ and Matrix Theory}{Adv.
Theor. Math. Phys.~{\bf 2} (1998) 51,} {\hh 9709220.}}

\lref\laroche{\paper {L.~Baulieu, C.~Laroche} {On Generalized Self-Duality
Equations Towards Supersymmetric   Quantum Field Theories Of
Forms}{\MPL{A13}{1998}{1115},}{\hh  9801014.}}

\lref\bsv{\paper {M.~Bershadsky, V.~Sadov, C.~Vafa} {$D$-Branes and
Topological Field Theories}{\NPB{463} {1996}{420},}{\hh9511222.}}

\lref\vafapuzz{\paper {C.~Vafa}{Puzzles at Large N}{}{\hh 9804172.}}

\lref\dvv{\paper {R.~Dijkgraaf, E.~Verlinde, H.~Verlinde} {Matrix String
Theory}{\NPB{500}{1997}{43},} {\hh9703030.}}

\lref\wynter{\paper {T.~Wynter}{Gauge Fields and Interactions in Matrix
String Theory}{\PLB{415}{1997}{349},}{\hh9709029.}}

\lref\kvh{\paper {I.~Kostov, P.~Vanhove}{Matrix String Partition
Functions}{}{\hh9809130.}}

\lref\ikkt{\paper {N.~Ishibashi, H.~Kawai, Y.~Kitazawa, A.~Tsuchiya} {A
Large $N$ Reduced Model as Superstring}{\NPB{498} {1997}{467},}{\hh
9612115.}}

\lref\ss{\paper {S.~Sethi, M.~Stern} {$D$-Brane Bound States
Redux}{\CMP{194}{1998} {675},}{\hh 9705046.}}

\lref\mns{\paper {G.~Moore, N.~Nekrasov, S.~Shatashvili} {$D$-particle Bound
States and Generalized Instantons}{} {\hh 9803265.}}

\lref\bsh{\paper {L.~Baulieu, S.~Shatashvili} {Duality from Topological
Symmetry}{} {\hh 9811198.}}

\lref\pawu{ {G.~Parisi, Y.S.~Wu} {}{ Sci. Sinica  {\bf 24} {(1981)} {484}.}}

\lref\coulomb{ {L.~Baulieu, D.~Zwanziger, }   {\it Renormalizable Non-Covariant
Gauges and Coulomb Gauge Limit}, {Nucl.Phys. B {\bf 548 } (1999) 527,} {\hh
9807024}.}

\lref\dan{ {D.~Zwanziger},  {}{Nucl. Phys. B {\bf   139}, (1978) {1}.}{}}

\lref\danzinn{  {J.~Zinn-Justin, D.~Zwanziger, } {}{Nucl. Phys. B  {\bf
295} (1988) {297}.}{}}

\lref\danlau{ {L.~Baulieu, D.~Zwanziger, } {\it Equivalence of Stochastic
Quantization and the Faddeev-Popov Ansatz,
  }{Nucl. Phys. B  {\bf 193 } (1981) {163}.}{}}

\lref\munoz{ { A.~Munoz Sudupe, R. F. Alvarez-Estrada, } {}
Phys. Lett. {\bf 164} (1985) 102; {} {\bf 166B} (1986) 186. }

\lref\okano{ { K.~Okano, } {}
Nucl. Phys. {\bf B289} (1987) 109; {} Prog. Theor. Phys.
suppl. {\bf 111} (1993) 203. }

\lref\singer{
 I.M. Singer, { Comm. of Math. Phys. {\bf 60} (1978) 7.}}

\lref\neu{ {H.~Neuberger,} {Phys. Lett. B {\bf 295}
(1987) {337}.}{}}

\lref\testa{ {M.~Testa,} {}{Phys. Lett. B {\bf 429}
(1998) {349}.}{}}

\lref\Martin{ L.~Baulieu and M. Schaden, {\it Gauge Group TQFT and Improved
Perturbative Yang--Mills Theory}, {  Int. Jour. Mod.  Phys. A {\bf  13}
(1998) 985},   hep-th/9601039.}

\lref\baugros{ {L.~Baulieu, B.~Grossman, } {\it A topological Interpretation
of  Stochastic Quantization} {Phys. Lett. B {\bf  212} {(1988)} {351}.}}

\lref\bautop{ {L.~Baulieu}{ \it Stochastic and Topological Field Theories},
{Phys. Lett. B {\bf   232} (1989) {479}}{}; {}{ \it Topological Field Theories
And Gauge Invariance in Stochastic Quantization}, {Int. Jour. Mod.  Phys. A
{\bf  6} (1991) {2793}.}{}}

\lref\samson{ {L.~Baulieu, S.L.~Shatashvili, { \it Duality from Topological
Symmetry}, {JHEP {\bf 9903} (1999) 011, hep-th/9811198.}}}{}

\lref\halpern{ {H.S.~Chan, M.B.~Halpern}{}, {Phys. Rev. D {\bf   33} (1986)
{540}.}}

\lref\yue{ {Yue-Yu}, {Phys. Rev. D {\bf   33} (1989) {540}.}}

\lref\neuberger{ {H.~Neuberger,} {\it Non-perturbative gauge Invariance},
{ Phys. Lett. B {\bf 175} (1986) {69}.}{}}

\lref\gribov{  {V.N.~Gribov,} {}{Nucl. Phys. B {\bf 139} (1978) {1}.}{}}

\lref\huffel{ {P.H.~Daamgard, H. Huffel},  {}{Phys. Rep. {\bf 152} (1987)
{227}.}{}}

\lref\creutz{ {M.~Creutz},  {\it Quarks, Gluons and  Lattices, }  Cambridge
University Press 1983, pp 101-107.}

\lref\zinn{ {J. ~Zinn-Justin, }  {Nucl. Phys. B {\bf  275} (1986) {135}.}}

\lref\shamir{  {Y.~Shamir,  } {\it Lattice Chiral Fermions
  }{ Nucl.  Phys.  Proc.  Suppl.  {\bf } 47 (1996) 212,  hep-lat/9509023;
V.~Furman, Y.~Shamir, Nucl.Phys. B {\bf 439 } (1995), hep-lat/9405004.}}

 \lref\kaplan{ {D.B.~Kaplan, }  {\it A Method for Simulating Chiral
Fermions on the Lattice,} Phys. Lett. B {\bf 288} (1992) 342; {\it Chiral
Fermions on the Lattice,}  {  Nucl. Phys. B, Proc. Suppl.  {\bf 30} (1993)
597.}}

\lref\neubergerr{ {H.~Neuberger, } {\it Chirality on the Lattice},
hep-lat/9808036.}

\lref\mandelstam{ {S. Mandelstam, } {} Phys. Rep. {\bf 23C}, 245
(1976).}

\lref\greensite{ {L. Del Debbio, M. Faber, J. Giedt, J. Greensite, and
S. Olejnik} {\it Detection of Center Vortices in the Lattice
Yang-Mills Vacuum}, Phys. Rev. {\bf D58} (1998) 094501.}

\lref\lueschweis{ {M. L\"{u}scher and P. Weisz, } {\it Quark
confinement and the bosonic string}, hep-lat/0207003.}

\lref\bali{ {G. S. Bali, } {\it Casimir scaling of SU(3) static
monopoles}, hep-lat/0006022.}

\lref\deldar{ {S. Deldar, } {\it Static SU(3) potentials for sources in
various representations}, hep-lat/9911008.}

\lref\simonov{ {V. I. Shevchenko, Yu. A. Simonov, } {\it Casimir
scaling as a test of QCD vacuum}, hep-ph/0001299.}

\lref\zbgr {L.~Baulieu and D. Zwanziger, {\it QCD $_4$ From a
Five-Dimensional Point of View},    hep-th/9909006.}

\lref\christlee{ {N. ~Christ and T. ~D. ~Lee,} {\it } {Phys. Rev.
{\bf  D22} (1980) {939}.}}

\lref\lattcoul{ {D. ~Zwanziger,} {\it Lattice Coulomb hamiltonian and
static color-Coulomb field,} {Nucl. Phys. B {\bf  485} (1997) {185}.}}

\lref\cnltcoul{ {D. ~Zwanziger,} {\it Continuum and Lattice
Coulomb-gauge hamiltonian,} {in {\it Cambridge 1997, Confinement,
duality, and non-perturbative aspects of QCD}, P. van Baal, Ed.}
{hep-th/9710157.}}

\lref\coul{ {D. ~Zwanziger,} {\it Renormalization in the Coulomb gauge and
order parameter for confinement in QCD,} {Nucl. Phys. B {\bf  518} (1998)
{237}.}}

\lref\recoul{ {L.~Baulieu and D. ~Zwanziger,} {\it Renormalizable non-covariant
gauges and Coulomb gauge limit,} {Nucl. Phys. B {\bf  548} (1999) {527}.}}

\lref\critical{ {D. ~Zwanziger,} {\it Critical Limit of Lattice Gauge Theory,} {Nucl.
Phys. B {\bf  378} (1992) {525}.}}

\lref\vanish{ {D. ~Zwanziger,} {\it Vanishing of zero-momentum lattice gluon
propagator and color confinement,} {Nucl. Phys. B {\bf  364} (1991) {127}.}}

\lref\robertson{ {D. G. Robertson, E. S. Swanson, A. P. Szczepaniak,
C.-R. Ji, S. R. Cotanch, } {\it  Renormalized Effective QCD Hamiltonian:
Gluonic Sector, }{Phys. Rev. D59 (1999) 074019.}}	

\lref\szcz{ {Adam Szczepaniak, Eric S. Swanson, Chueng-Ryong Ji, Stephen
R. Cotanch, } {\it  Glueball Spectroscopy in a Relativistic Many-Body
Approach to Hadron Structure, }{Phys. Rev. Lett. 76 (1996) 2011-2014.}}

\lref\szczsw{ {A. P. Szczepaniak, E. S. Swanson, } {\it  Coulomb gauge
QCD, confinement and the constituent representation,
}{ Phys.Rev. {\bf D65}:025012 (2002).}}		

\lref\cuzwsc{ {Attilio Cucchieri, Daniel Zwanziger, } {\it  Static
Color-Coulomb Force, }{Phys. Rev. Lett. 78 (1997) 3814-3817.}}	

\lref\ZZ{ {Ismail Zahed, Daniel Zwanziger, } {\it  Zero Color Magnetization in
QCD Matter, }{Phys. Rev. D61 (2000) 037501.}}	

\lref\cuzwns{ {Attilio Cucchieri, Daniel Zwanziger, } {\it  Numerical
study of gluon propagator and confinement scenario in minimal Coulomb
gauge, } {Phys. Rev. D65 (2001) 014001.}}	

\lref\rengrcoul{ {Attilio Cucchieri, Daniel Zwanziger, } {\it 
Renormalization-group calculation of the color-Coulomb, }{Phys. Rev.
D65 (2001) 014002.}}	

\lref\ccoulpot{ {Attilio Cucchieri, Daniel Zwanziger, } 
{\it Gluon propagator and confinement scenario in Coulomb gauge,}
{hep-lat/0209068.}}	

\lref\pesschro{ {Michael Peskin, Daniel Schroeder, } {\it  An
Introduction to field theory, }{Perseus Books (1995) p. 593.}}	

\lref\doust{ {R. N. Doust, } {\it  Ann. of Phys., }{177 (1987) 169.}}	

\lref\taylor{ {J. C. Taylor, } {\it Physical and Non-standard Gauges,}
{Proc., Vienna, Austria, 1989, ed. P. Gaigg, W. Kummer and M.
Schweda (Springer, Berlin, 1990) p. 137.}}	

\lref\tdlee{ {T. D. Lee, } {\it  Particle physics and
introduction to field theory, }{Harwood (1981) p. 455.}}


\nfig\compar{
Horizontal lines correspond to instantaneous propagators $1/\vk^2$. 
Curved lines correspond to non-instantaneous propagators 
$1/(\vk^2 + k_4^2)$.  Diagram 1a is a contribution to $V_0$.  Diagram
1b is a contribution to $P_0$.}
 


\Title{\vbox
{\baselineskip 10pt
\hbox{hep-th/0209105}
\hbox{NYU-TH-PH-20.8.00}
\hbox{BI-TP 2000/19}
 \hbox{   }
}}
{\vbox{\vskip -30 true pt
\centerline{ No confinement without Coulomb confinement}
\medskip
\vskip4pt }}
\centerline{
{\bf  Daniel Zwanziger}$^{ \dag}$}
\centerline{
Daniel.Zwanziger@nyu.edu}
\vskip 0.5cm

\centerline{\it $^{\dag}$   Physics Department, New York University,
New-York,  NY 10003,  USA}

\medskip
\vskip  1cm
\noindent

\def\vx{\vec{x}}
\def\vk{\vec{k}}

	We compare the physical potential $V_D(R)$ of an external
quark-antiquark pair in the representation $D$ of SU(N), to the
color-Coulomb potential $V_{\rm coul}(R)$ which is the instantaneous
part of the 44-component of the gluon propagator in Coulomb gauge 
$D_{44}(\vx,t) = V_{\rm coul}(|\vx|) \d(t)$ + (non-instantaneous).  We
show that if $V_D(R)$ is confining, 
$\lim_{R \to \infty}V_D(R) = + \infty$, then the inequality $V_D(R)
\leq - C_DV_{\rm coul}(R)$ holds asymptotically at large $R$, where $C_D
> 0$ is the Casimir in the representation $D$.  This implies that $ -
V_{\rm coul}(R)$ is also confining.

\Date{\ }

\def\a{\alpha}
\def\b{\beta}
\def\d{\delta}
\def\c{\gamma}
\def\m{\mu}

\def\r{\rho}
\def\s{\sigma}
\def\l{\lambda}
\def\L{\Lambda}

\def\F{\Phi}

\def\vx{\vec{x}}
\def\vy{\vec{y}}

\newsec{Introduction}

	The problem of confinement of color charge has been with us for a long
time.  There are many approaches to this problem such as 
dual Meissner effect by monopole condensation~\mandelstam, effective
string theory~\lueschweis, center dominance~\greensite, and
color-Coulomb potential~\coul.  One seeks to choose variables so that
the most important degrees of freedom have a simple expression.  For
this purpose gauge fixing can be a useful technique.

	Confinement is most commonly characterized by the behavior of
$V_D(R)$, the gauge-invariant potential energy between an external quark
pair at separation $R$ in the representation $D$ of the gauge structure
group SU(N). It may found from a rectangular Wilson loop 
$W_D(R, T)$ in the representation $D$, but for our purposes it is more
convenient to obtain it from the correlator, 
$\langle P_D(\vx)P_D^*(\vy) \rangle$, of a pair of Polyakov or thermal
Wilson loops at $\vx$ and $\vy$  in the representation~$D$, on a
Euclidean lattice of period $T$ in the 4-direction.  The Polyakov loop
is the lattice analog of the continuum expression  
$P_D(\vx) = {\rm tr}[P\exp(\int_0^T A_{D,4}(\vx,t) dt]$, where
$A_{D,\m} \equiv A_\m^a t_D^a$, and the 
$t_D^a$ satisfy the Lie algrebra commutation relations
$[t_D^a, t_D^b] = f^{abc}t_D^c$ in the representation~$D$.  As
discussed recently~\lueschweis, this correlator
has the expansion
\eqn\pol{\eqalign{
\langle P_D(\vx)P_D^*(\vy) \rangle
= \sum_{n=0}^\infty \exp(-E_{n, \vx, \vy}T),
}}
where the $E_{n, \vx, \vy}$ are the eigenvalues,
$H \Psi_n = E_{n,\vx, \vy} \Psi_n$, of the lattice QCD 
hamiltonian $H$, specified below, that includes an
external quark and anti-quark at $\vx$ and $\vy$ in the
representation~$D$.  In the large-$T$ limit, the sum is dominated by the
first term, with lowest energy eigenvalue,
$E_{0,\vx,\vy}$.  It is rotationally symmetric in the continuum
limit, and we identify the physical quark-antiquark
potential $V_D(R)$ with this energy eigenvalue,
after separation of divergences, 
\eqn\vd{\eqalign{
E_{0,|\vx_1-\vx_2|} = E_0 + \Delta +  V_D(|\vx - \vy|).
}}  
Here $E_0$ is the energy of the vacuum state in the absence of an
external quark pair, and $\Delta$ is the diverging self-energy of the
external quarks.  According to the Wilson confinement criterion, which is
expected to hold in pure gluodynamics without dynamical quarks, but with
external quarks in the fundamental representation, 
$V_F(R)$ diverges linearly at large
$R$, $V_F(R) \sim \s R$ where $\s$ is the conventional string
tension.

In the Coulomb gauge, there is a simple scenario~\coul\ that
attributes confinement of color charge to the long range of the
the color-Coulomb potential, $V_{\rm coul}(R)$.  This quantity
characterizes the instantaneous part of the 44-component of the gluon
propagator 
$\langle A_4^a(\vx,t)A_4^b(0,0) \rangle = D_{44}(\vx,t) \d^{ab} 
= V_{\rm coul}(|\vx|) \ \d(t) \ \d^{ab}$ + (non-instantaneous).  Since
$A_4$ couples universally to the color-charge, this can
account for confinement of color-charge, provided that $V_{\rm coul}(R)$
is indeed long range.  A remarkable feature of the Coulomb gauge in
QCD, a property not shared by any Lorentz gauge, is that
$A_4 = g_0 A_{4}^{(0)} = g_r A_{4}^{(r)}$ 
is a renormalization-group invariant
\coul, \coulomb.  Here $g_0$ and
$A_{4}^{(0)}$, and $g_r$ and $A_{4}^{(r)}$ are,
respectively, the unrenormalized and renormalized charges and
perturbative gauge
connections.  This means that $D_{44}$, and hence also its
instantaneous part $V_{\rm coul}(R)$, is independent of both the cut-off
$\L$ and the renormalization mass $\m$.  This property allows the
fundamental QCD quantity, $V_{\rm coul}(R)$, the instantaneous part of
the gluon propagator, to be identified with the phenomenological QCD
potential \szcz\ and \szczsw.   Its fourier transform,
$\tilde{V}_{\rm coul}(|\vec{k}|)$, provides a
convenient definition of the running coupling constant, 
$\a_s(|\vec{k}|/\L_{\rm coul}) 
= g_{\rm coul}^2(|\vec{k}|)/(4\pi)
= \vec{k}^2\tilde{V}_{\rm coul}(|\vec{k}|)/(4\pi x_0)$,
where $x_0 = 12N/(11N - N_f)$, $\L_{\rm coul}$ is a finite
QCD mass scale, and $N_f$ is the number of quark flavors~\rengrcoul. 
The result obtained here means that if the Wilson criterion for
confinement is satisfied, then, with this definition, the running
coupling constant $\a_s(|\vec{k}|/\L_{\rm coul})$ diverges
in the infrared like $1/\vec{k}^2$, a clear manifestation of infrared
slavery.

 We shall show that a necessary condition for
confinement according to the Wilson criterion is that the instantaneous
color-Coulomb potential be confining.  In symbols: if
$\lim_{R \to \infty}V_D(R) = +\infty$, then
$V_D(R) \leq	- \ C_D \ V_{\rm coul}(R)$
holds asymptotically at large $R$.  The minus sign occurs because
antiquark has opposite charge to quark.  Here 
$C_D = - \sum_a(t_D^a)^2 > 0$ is the (positive) value of the Casimir
invariant, and in the fundamental representation 
$C_F = (N^2-1)/(2N)$.  A considerable simplification is hereby achieved
because $V_D(R)$ is defined by means of a
path-ordered exponential that involves gluon $n$-point functions of all
orders, whereas $V_{\rm coul}(R)$ is defined in
terms of the gluon 2-point function, $D_{44}$.  We also note the striking
numerical result~\bali, \deldar\ that $V_D(R)$ exhibits Casimir
scaling, $V_D(R) = (C_D/C_F)V_F(R)$ quite accurately at least in a
rather large range of $r$ and 8 representations $D$.  This suggests that
the above bound may be saturated in this range, for this would explain
Casimir scaling, that is not easy to understand otherwise~\simonov.
The result also makes it imperative to extend present programs
to calculate $V_{\rm coul}(R)$ numerically~\ccoulpot, and
analytically from first principles~\szczsw, and to derive
phenomenological quantities from it~\szcz. 

\newsec{Lattice Coulomb-gauge QCD hamiltonian} 

	The energy $E_{0,|\vx_1-\vx_2|}$ is of course gauge invariant, and the
lattice QCD hamiltonian $H$ may be chosen in any gauge.  Its most
familiar form is in the temporal gauge $U_4 = 1$, corresponding to
$A_4 = 0$,  
\eqn\latham{\eqalign{
H_{\rm temp} = g_0^2(2a)^{-1} 
\sum_{\vx,i} {\cal E}_{\vx,i}^2 
+ 2(g_0^2a)^{-1}\sum_p {\rm Re \ Tr}U_p,
}}
where $\sum_p$ is the sum over all spatial plaquettes $p$ (on a single
time-slice).  Here ${\cal E}_{\vx,i}$ is the color-electric field
operator  that satisfies
$[{\cal E}_{\vx,i}^a,U_{\vy,j}] = i t^a U_{\vy,j} \d_{\vx,\vy} \d_{ij}$
and
$[{\cal E}_{\vx,i}^a, {\cal E}_{\vy,j}^b ] = 
-i f^{abc} \d_{\vx,\vy} \d_{ij}{\cal E}_{\vx,i}^c$.  
We place an external quark at $\vx_1$ in the representation $D$, and an
external anti-quark at $\vx_2$ in the representation $D^*$, with color
vectors that act on the first and second indices of the wave-functional
$\Psi_{\a \b}(U)$ according to
$(\l_1^a\Psi)_{\a \b} = (\l_D^a)_{\a \c}\Psi_{\c \b}$
and
$(\l_2^a\Psi)_{\a \b} = - (\l_D^a)_{\b \c}^*\Psi_{\a \c}$,
where $\l_D^a = it_D^a$.  In the temporal gauge, the color charges of
the external quarks  do not appear in the hamiltonian $H_{\rm temp}$,
but rather in  the subsidiary condition $G^a(\vx) \Psi = 0$.
This is an expression of Gauss's law, for $G^a(\vx)$ is a precise
lattice analog of the continuum Gauss's law operator
$G^a(\vx) = - (\vec{D}\cdot \vec{E})^a(\vx) + \r_{\rm qu}^a(\vx)$, where
$E_i^a(\vx) = i{ {\d} \over {\d A_i^a(\vx)} }$,
$D_i^{ac} = \d^{ac}\p_i + f^{abc}A_i^b$ is the gauge-covariant
derivative, and 
$\r_{\rm qu}^a(\vx) = \l_1^a \d(\vx-\vx_1) + \l_2^a \d(\vx-\vx_2)$
is the color-charge density of the external quarks.  In the temporal
gauge $G^a(\vx)$ is the generator of 3-dimensionally local gauge
transformations of the quark and gluon variables, satisfying
$[G^a(\vx), G^b(\vy)] = i\d(\vx - \vy) f^{abc} G^c(\vx)$, and the
subsidiary condition is the statement of gauge invariance of the wave
functional.  

	One would expect that Gauss's law is essential for confinement, and
the lattice Coulomb hamiltonian $H_{\rm coul}$ \lattcoul\ may be
derived from $H_{\rm temp}$ by solving Gauss's law as subsidiary
condition \cnltcoul. For our purposes the resulting lattice Coulomb
hamiltonian has the same structure to the continuum Coulomb hamiltonian
\christlee.  To simplify the exposition, we shall use continuum
language, but it is understood that this is short-hand for the correct
lattice kinematics, and divergences are controlled by use of the lattice
Coulomb hamiltonian, as will be made clear.

To get to
the Coulomb gauge from the temporal gauge, one integrates out the gauge
degrees of freedom using the Faddeev-Popov formula in all
gauge-invariant matrix elements.  In particular for the hamiltonian, one
obtains $H_{\rm coul}$, defined by its matrix elements
\eqn\christtd{\eqalign{
(\Psi_1, H_{\rm coul} \Psi_2) = 
\int_\Lambda dA^{\rm tr} \det M \ (1/2) \int d^3x \
 [g_0^2(E_i^a \Psi_1)^* E_i^a \Psi_2 
+ g_0^{-2}\Psi_1^* \vec{B}^2 \Psi_2],
}}
there the wave-functionals
$\Psi_{\a\b}(A^{\rm tr})$ depend only on 3-dimensionally transverse
continuum configurations
$\p_i A_i^{\rm tr} = 0$,
and a contraction on color indices is understood.
The color-magnetic field is given by 
$B_1^a = \p_2 A_3^{{\rm tr},a} - \p_3 A_2^{{\rm tr},a}
 + f^{abc}A_2^{{\rm tr},b}A_3^{{\rm tr},c}$, etc.,
and the color-electric field by   
$E_i^a = E_i^{{\rm tr},a} - \p_i \f^a$, where 
$E_i^{{\rm tr},a} = i { {\d} \over {\d A_i^{{\rm tr},a} } } $,
and $\f^a(\vx)$ is the color-Coulomb potential operator.
In this matrix element,~$\f^a(\vx)$ acts directly on
the wave functional~$\Psi$.  The definition of $H_{\rm coul}$ is
completed by specifying that 
$\f^a(\vx) \Psi \equiv (M^{-1}\r_{\rm phys})^a(\vx)\Psi$,
which expresses $\f^a(\vx)\Psi$ in terms $\r_{\rm qu}$ and transverse
gluon variables only.  This is the solution of the subsidiary condition 
$G^a(\vx)\Psi = 0$, or
$M^{ac}(A^{\rm tr}) \f^c \Psi = \r_{\rm phys}^a \Psi$.
Here 
$M^{ac}(A^{\rm tr}) \equiv - D_i^{ac}(A^{\rm tr})\p_i
= - \p_iD_i^{ac}(A^{\rm tr}) 
= - \p^2 \d^{ac} - f^{abc} A_i^{{\rm tr},b}\p_i$ 
is the 3-dimensional Faddeev-Popov operator, and
$\r_{\rm phys}^a 
   \equiv - f^{abc} A_i^{{\rm tr},b} E_i^{{\rm tr},c} + \r_{\rm qu}^a$
is the color-charge density of the external quarks plus the
color-charge of the dynamical gluon degrees of freedom only.  The
associated color charge, 
$Q^a = \int d^3x \ \r_{\rm phys}^a(\vx)$ may be identified with the
physical color charge, for it generates global gauge
transformations on all variables
$[Q^a, A_i^{{\rm tr},b}] = if^{abc}A_i^{{\rm tr},c}$ etc., and
satisfies  $[Q^a, Q^b] = i f^{abc} Q^c$.  The second term in $M$ is
characteristic of non-Abelian gauge theory.  It is responsible for
anti-screening because, for typical
configurations, this term produces a small denominator in $M^{-1}$.  The
subscript $\L$ on the integral $\int_\Lambda dA^{\rm tr}$  means that a
region that includes only one Gribov copy is integrated over.   
This may be chosen as in the minimal Coulomb gauge, but the
proof does not depend on the particular way this is chosen.

\newsec{Bound on $V_D(R)$ from trial wave function}	

 The energy $E_{|\vx_1-\vx_2|} \equiv (\Psi, H_{\rm coul} \Psi)$
of any trial wave function $\Psi$ provides an upper bound on the
ground-state energy, 
$E_{0,|\vx_1-\vx_2|} \leq E_{|\vx_1-\vx_2|}$.  As trial function
we take the product wave function,
$\Psi_{\a \b}(A^{\rm tr}) = N_D^{-1/2}\d_{\a \b} \ \F_0(A^{\rm tr})$. 
Here $\F_0(A^{\rm tr})$ is the exact wave functional of the vacuum state
in the absence of external quarks, and  
$N_D^{-1/2}\d_{\a\b}$, where $N_D$
is the dimension of representation $D$, is the external
quark-pair state of total color-charge
zero, $(\l_1 + \l_2)^a \Psi = 0$.  
The Coulomb hamiltonian has the decomposition
$H_{\rm coul} = H_{\rm gl} + H_{\rm gl, qu} + H_{\rm qu,qu}$, 
that follows from the decomposition of the color-charge density
$\r_{\rm phys}^a 
   = - f^{abc} A_i^{{\rm tr},b} E_i^{{\rm tr},c} + \r_{\rm qu}^a$ 
in $\f \ \Psi = M^{-1}\r_{\rm phys} \ \Psi$.  Here
$H_{\rm gl}$ is the Coulomb hamiltonian in the absence of external
quarks,
$H_{\rm gl, qu}$ is linear in $\r_{\rm qu}$, and 
$H_{\rm qu,qu} = (1/2)\int d^3x \ (\p_i M^{-1} \r_{\rm qu})^2(\vx)$.
In the last expression there is no ordering problem because $\r_{\rm
qu}$ commutes with $A^{\rm tr}$.  By definition of $\Phi_0$  we have
$H_{\rm gl} \Phi_0 = E_0 \Phi_0$,
where $E_0$ is the vacuum energy in the absence of
external quarks.  From
$(\Psi, H_{\rm gl}\Psi) = E_0$,
and
$(\Psi, H_{\rm gl, qu}\Psi) = 0$,
we get for the trial energy, 
\eqn\trial{\eqalign{
E_{|\vx_1 - \vx_2|} & = E_0 + \Delta'
	- \ C_D 
(\F_0, [M^{-1}(-\p^2)M^{-1}]_{\vx_1,\vx_2}^{a\ \ a}\F_0) 
}}
(no sum on $a$), where we have used 
$\langle\l_1^a \l_2^b \rangle 
= (N^2-1)^{-1}\d^{ab}\langle\l_1^c \l_2^c \rangle
= -(N^2-1)^{-1}\d^{ab}C_D$.  
Here $\Delta'$ is another self energy of the external
quarks that is independent of $\vx_1$ and $\vx_2$.  The inequality 
$E_{0,|\vx_1 - \vx_2|} \leq E_{|\vx_1 - \vx_2|}$
reads
$\Delta(\L) + V_D(R, \L) \leq \Delta'(\L) - \ C_D 
(\Phi_0, [M^{-1}(-\p^2) M^{-1}]_{\vx_1,\vx_2}^{a\ \ a} \ \Phi_0)$.  We
have cancelled the vacuum energy $E_0$ that diverges with the volume of
space and, having done so, we may take the volume of space to infinity,
keeping the ultraviolet cut-off $\L = a^{-1}$ in place, where $a$ is the
lattice spacing.  Here
$\Delta(\L)$ and $\Delta'(\L)$ are self-energies of the external
quarks.  The formula \rengrcoul\ for the color-Coulomb potential,
$V_{\rm coul}(|\vx_1 - \vx_2|) \d^{ab} 
	= (\Phi_0, [M^{-1}(-\p^2) M^{-1}]_{\vx_1,\vx_2}^{a\ \ b} \ \Phi_0)$,
allows us to write the inequality as 
$\Delta(\L) + V_D(R, \L) \leq \Delta'(\L) - \ C_D V_{\rm coul}(R, \L)$.

	We have inserted a dependence on the cut-off $\L$ in 
$V_D(R, \L)$ and  $V_{\rm coul}(R, \L)$, because these are
lattice quantities that depend on the lattice spacing, $a = \L^{-1}$. 
However, having separated out the self-energies,
both the quark potential $V_D(R, \L)$ and the lattice color-Coulomb
potential $V_{\rm coul}(R, \L)$
have  finite, $\L$-independent continuum limits,
$\lim_{\L \to \infty}V_D(R, \L) = V_D(R)$ and
$\lim_{\L \to \infty}V_{\rm coul}(R, \L) = V_{\rm coul}(R)$.
For $V_D(R)$ is a physical energy, and $V_{\rm coul}(R)$
is a renormalization-group invariant, as noted in the
Introduction.  If $V_D(R, \L)$ is confining,
$\lim_{R \to \infty} V_D(R, \L) = +\infty$, then, for sufficiently
large $R$, the self-energies 
$\Delta(\L)$ and $\Delta'(\L)$ are negligible compared to 
$V_D(R, \L)$, and the inequality 
$V_D(R, \L) \leq - \ C_D \ V_{\rm coul}(R, \L)$ holds asymptotically at
large~$R$, for finite cut-off $\L$.  This bound also holds in
the continuum limit, because dimensional and renormalization-group
considerations tell us that the terms that vanish as $\L \to \infty$ are
of relative order $1/(\L R)^n$, where $n$ is positive, so they also
vanish asymptotically at large~$R$.\foot{This condition is necessary,
as shown by the following counter-example.  Take 
$V_D(R, \L) = \s R$, and $V_{\rm coul}(R, \L) = c/R + m^4 R^2/\L$, with
self-energies
$\D(\L) = a \ \L$, and $\D'(\L) = (a + 1) \L$.
The inequality at finite $\L$ reads
$\s R \leq \L + c/R +  m^4 R^2/\L = 2m^2R + c/R + (\L - m^2 R)^2/\L$. 
It is satisfied for all finite $\L$ and $R$, provided that
$\s \leq 2m^2$ and $c \geq 0$.  But the continuum limit
of $V_{\rm coul}(R)$ is $c/R$, 
and $\s R < c/R$ is {\it not} satisfied at large $R$. }
We conclude that, if $V_D(R)$ is confining, 
$\lim_{R \to \infty} V_D(R) = +\infty$,
then in the continuum limit, the inequality, 
$V_D(R) \leq - \ C_D \ V_{\rm coul}(R)$, holds asymptotically at
large~$R$, as asserted.

\newsec{Conclusion}

The bound implies that if the potential between external quarks in the
fundamental representation increases linearly at large $R$, 
$V_F(r) \sim \s R$, where $\s$ is the standard string tension, then
the color-Coulomb potential
$V_{\rm coul}(R)$ increases at least linearly at large $R$, and moreover
if its increase is also linear, 
$ - V_{\rm coul}(R) \sim \s_{\rm coul}R$, 
as has been conjectured \coul, where $\s_{\rm coul}$ is a string
tension that characterizes 
$V_{\rm coul}(R)$, then the conventional string tension satisfies the
bound $\s \leq (N^2-1)/(2N)\s_{\rm coul}$.

	What has been learned about QCD dynamics?  We have found that if the
Wilson confinement criterion holds, then $V_{\rm coul}(R)$, the
instantaneous part of the gluon propagator $D_{44}$ in Coulomb gauge,
is confining.  Moreover, from 
$V_{\rm coul}(|\vx-\vy|)
= \langle[M^{-1}(-\p^2)M^{-1}]_{\vx,\vy}\rangle$, 
this can happen only if the Faddeev-Popov or ghost Green function,
$[M^{-1}(A^{\rm tr})]_{\vx,\vy}$, is long-range for configurations
$A^{\rm tr}$ that dominate the functional integral.  This confirms
the confinement scenario originally proposed by Gribov~\gribov, and
advocated by the author~\coul.  [The scenario reads, in brief, that
in the minimal Coulomb gauge, configurations are restricted to the
Gribov region, where the Faddeev-Popov operator is positive, 
$M(A^{\rm tr})>0$.  The boundary of this region occurs where the
lowest eigenvalue of $M(A^{\rm tr})$ vanishes, $\l_0(A^{\rm tr}) = 0$. 
Moreover the dimension $n$ of configuration space is very large, being
of the order of the volume $V$ of the lattice.  Entropy favors 
a population highly concentrated close to this boundary, where
$\l_0(A^{\rm tr})$ is small, for the same reason that, in a space of
very high dimension $n$, the density of a ball $r < r_0$ is very sharply
peaked near $r_0$, being given by~$r^{n-1}dr$.  Consequently, for the
configurations that dominate the functional integral, 
$M^{-1}(A^{\rm tr})$ is enhanced, and thus also~$V_{\rm coul}(R)$.]

	To simplify the exposition, we considered gluodynamics without
dynamical quarks, but the proof holds if they are included.  However if
dynamical quarks are present in the fundamental representation~F,
as occurs in nature, then the physical potential $V_F(R)$ between
external quarks does not manifest confinement.  For at some radius $R_b$
the string breaks by polarization of sea quarks from the vacuum, and
for $R > R_b$, $V_F(R)$ represents a residual potential between a pair
of mesons, analogous to the van der Waals potential.  In this case the
bound obtained here does not imply that $V_{\rm coul}(R)$ is confining. 
Nevertheless, according to the confinement scenario in Coulomb gauge,
$V_{\rm coul}(R)$ is a fundamental quantity that remains linearly
rising even when $V_F(R)$ is not.  It is precisely the linear rise 
of $V_{\rm coul}(R)$ that causes string-breaking, by making it
energetically preferable to polarize sea-quarks from the vacuum.

\vskip 3mm

  \centerline{\bf Acknowledgments}

This research was partially supported by the National
Science Foundation under grant PHY-9900769.  The author recalls with
pleasure valuable conversations with Lorenz von Smekal, and
 stimulating discussions and correspondence with Jeff Greensite
that were the origin of this investigation.

\vskip 2cm

\footatend\vfill\supereject\immediate\closeout\rfile\writestoppt
\baselineskip=14pt\centerline{{\bf References}}\bigskip{\frenchspacing%
\parindent=20pt\escapechar=` \input refs.tmp\vfill\eject}\nonfrenchspacing



\bye